\documentclass[aps,showpacs,nofootinbib,superscriptaddress,twocolumn,floatfix]{revtex4-1}

\usepackage{graphicx}
\usepackage{amssymb}
\usepackage{amsmath}
\usepackage{bm}
\usepackage{slashed}
\usepackage{datetime}
\usepackage{mciteplus}

\begin{document}
\allowdisplaybreaks[2]

\title{Quark angular momentum and the Sivers asymmetry}

\author{Alessandro Bacchetta}
\email{alessandro.bacchetta@unipv.it}
\affiliation{Dipartimento di Fisica, Universit\`a di Pavia, via Bassi 6, I-27100 Pavia, Italy}
\affiliation{INFN Sezione di Pavia, via Bassi 6, I-27100 Pavia, Italy}

\author{Marco Radici}
\email{marco.radici@pv.infn.it}
\affiliation{INFN Sezione di Pavia, via Bassi 6, I-27100 Pavia, Italy}

\begin{abstract}
The determination of quark angular momentum requires the knowledge of the generalized parton 
distribution $E$ in the forward limit. We assume a connection between this function and the Sivers 
transverse-momentum distribution, based on model calculations and theoretical considerations. Using 
this assumption, we show that it is possible to fit at the same time nucleon anomalous magnetic moments and 
semi-inclusive single-spin asymmetries. This imposes additional constraints on the Sivers function and 
opens a plausible way to quantifying the quark angular momentum.
\end{abstract}

\date{\today, \currenttime}

\maketitle


\section{Introduction}

This talk describes the results of a recent work~\cite{Bacchetta:2011gx}, where we proposed a nonstandard 
way to constrain the angular momentum $J^a$ of a (anti)quark with flavor $a$. In order to do this, we adopted an
assumption~\cite{Burkardt:2002ks}, motivated by model calculations and theoretical considerations, that connects
$J^a$ to the Sivers transverse-momentum distribution (TMD)~\cite{Sivers:1990cc} measured in semi-inclusive DIS (SIDIS). We showed that this assumption is compatible with existing data, and allows us to derive estimates of 
$J^a$ in fair agreement with other ``standard'' extractions. 

The total angular momentum of a parton $a$ (with $a = q, \bar{q}$) at some scale $Q^2$ can be computed as a specific moment of generalized parton distribution functions (GPD)~\cite{Ji:1997ek} 
\begin{equation}
2 J^{a} (Q^2)=  \int_0^1 dx \, x\, \bigl(H^{a}(x,0,0; Q^2)+E^{a}(x,0,0; Q^2) \bigr).
\label{e:Jdef}
\end{equation} 
The GPD $H^a(x,0,0; Q^2)$ corresponds to the familiar collinear parton distribution function (PDF)
$f_1^{a}(x; Q^2)$. The forward limit of the GPD $E^a$ does not correspond to any collinear PDF. It is possible to probe the function $E^a$ in experiments, but never in the forward limit~\cite{Kumericki:2007sa}. Assumptions are eventually necessary to constrain $E^a(x,0,0; Q^2)$. This makes the estimate of  $J^a$ particularly challenging. The only 
model-independent constraint is the scale-independent sum rule   
\begin{equation}
\sum_q \  e_{q_v} \int_0^1 dx \, E^{q_v}(x,0,0) = \kappa , 
\label{eq:kappa}
\end{equation}
where $E^{q_v}=E^q-E^{\bar{q}}$ and $\kappa$ denotes the anomalous magnetic moment of the parent nucleon. 

Denoting the Sivers function by $f_{1T}^{\perp a}$, we propose this simple relation at a scale $Q_L$,  
 \begin{equation}
 f_{1T}^{\perp (0) a}(x; Q_L^2) = -L(x)\, E^a (x,0,0; Q_L^2) ,
\label{e:EtoSivers0}
\end{equation}  
where $f_{1T}^{\perp(0) a}$ is the integral of the Sivers function over transverse momentum. This assumption is inspired by theoretical considerations~\cite{Burkardt:2002ks} and by results of spectator 
models~\cite{Burkardt:2003je,Lu:2006kt,Meissner:2007rx,Bacchetta:2008af}.  $L(x)$ is a flavor-independent function, representing the effect of the QCD interaction of the outgoing quark with the rest of the nucleon. The name ``lensing function'' has been proposed by Burkardt to denote $L(x)$~\cite{Burkardt:2003uw}. Computations of the lensing function beyond the single-gluon approximation have been proposed in  Ref.~\cite{Gamberg:2009uk}.  It is likely that in more complex models the above relation is not preserved. Nevertheless, it is useful and interesting to speculate on the consequences of this simple assumption. 

The advantage of adopting the Ansatz of Eq.~\eqref{e:EtoSivers0} is twofold: first, the value of the anomalous magnetic moment fixes the integral of the GPD $E$ and allows us to constrain the valence Sivers function also outside the region where SIDIS data are available; second, our Ansatz allows us to obtain flavor-decomposed information on the $x$-dependence of  the GPD $E$ and, ultimately, on quark and antiquark total angular momentum. This is an example of how assuming model-inspired connections between GPD and TMD can lead to
powerful outcomes.

\section{Fitting the data}

To analyze SIDIS data, we use the same assumptions adopted in 
Refs.~\cite{Arnold:2008ap,Anselmino:2011gs}: we neglect the effect of TMD evolution, which has been studied only recently~\cite{Aybat:2011ge,Aybat:2011ta,Anselmino:2012aa}; we assume a flavor- and 
scale-independent Gaussian transverse-momentum distribution for all involved TMDs, and we include the effect of the standard DGLAP evolution only in the collinear part of the parametrizing functions. 

Neglecting $c, b, t$ flavors, we parametrize the Sivers function in the following way: 
\begin{align}
f_{1T}^{\perp q_v}(x,k_{\perp}^2;Q_0^2) &=  C^{q_v} \frac{\sqrt{2 e} M M_1}{\pi M_1^2 \langle k_{\perp}^2 \rangle }\, 
(1-x) f_1^{q_v}(x; Q_0^2) \nonumber \\
&\; \times e^{-k_{\perp}^2/M_1^2}  e^{-k_{\perp}^2/\langle k_{\perp}^2 \rangle} \, 
\frac{1- x / \alpha^{q_v}}{|\alpha^{q_v}-1|}\, .
\label{e:gauss}
\end{align}
For $\bar{q}$, we use a similar function, excluding the last term. We used 
$\langle k_{\perp}^2 \rangle= 0.14$ GeV$^2$. $M_1$ is a free parameter that determines the 
transverse-momentum width. We imposed constraints on the parameters $C^a$ in order to respect the positivity bound for the Sivers function~\cite{Bacchetta:1999kz}. We multiply the unpolarized PDF by $(1-x)$ to respect the predicted high-$x$ behavior of the Sivers function~\cite{Brodsky:2006hj}. We introduce the free parameter  
$\alpha^{q_v}$ to allow for the presence of a node at  $x=\alpha^{q_v}$, as suggested in
Refs.~\cite{Bacchetta:2008af,Bacchetta:2010si,Kang:2012xf,Boer:2011fx}. 

For the lensing function we use the following Ansatz
\begin{equation}
L(x) = \frac{K}{(1-x)^{\eta}} .
\label{e:lensing}
\end{equation} 
The choice of this form is guided by model 
calculations~\cite{Burkardt:2003je,Lu:2006kt,Meissner:2007rx,Bacchetta:2008af,Bacchetta:2010si}, by the 
large-$x$ limit of the GPD $E$~\cite{Brodsky:2006hj}, and by the phenomenological analysis of the GPD $E$ proposed in Ref.~\cite{Guidal:2004nd}. We checked {\it a posteriori} that there is no violation of the positivity bound on the GPD $E^{q_v}$. 

We performed a combined $\chi^2$ fit to the nucleon anomalous magnetic moments (for our present purposes, we take $\kappa^p = 1.793 \pm 0.001, \; \kappa^n = -1.913 \pm 0.001$) and the Sivers asymmetry with identified hadrons from Refs.~\cite{Airapetian:2009ti,Alekseev:2008dn,Qian:2011py}.

We set the gluon Sivers function to zero (its influence through evolution is anyway limited) and we chose 
$Q_0 = Q_L=1$ GeV. We fixed $\alpha^{d_v,s_v}=0$ (no nodes in the valence down and strange Sivers functions).
We explored several scenarios characterized by different choices of the parameters related to the strange quark. 
In all cases, we obtained very good values of $\chi^2$ per degree of freedom ($\chi^2$/dof), around 1.34.

All fits lead to a negative Sivers function for $u_v$ and large and positive for $d_v$, in agreement with previous  
studies~\cite{Vogelsang:2005cs,Anselmino:2008sga,Arnold:2008ap,Anselmino:2011gs}. The data are compatible with vanishing sea-quark contributions (with large uncertainties). However, in the $x$ range where data exist, large Sivers functions for $\bar{u}$ and $\bar{d}$ are excluded, as well as large and negative for $\bar{s}$. The Sivers function for $s_v$ is essentially unconstrained. The parameter $M_1$ is quite stable around 0.34 GeV, as well as
the strength of the lensing function $K$ around 1.86 GeV. The parameter $\eta$ is typically around 0.4. It turns out  
$\alpha^{u_v} \approx 0.78$, so there is little room for a node in the up Sivers function, also because of the constraint imposed by the anomalous magnetic moments.

\begin{figure}[h]
\begin{center}
\includegraphics[width=6cm]{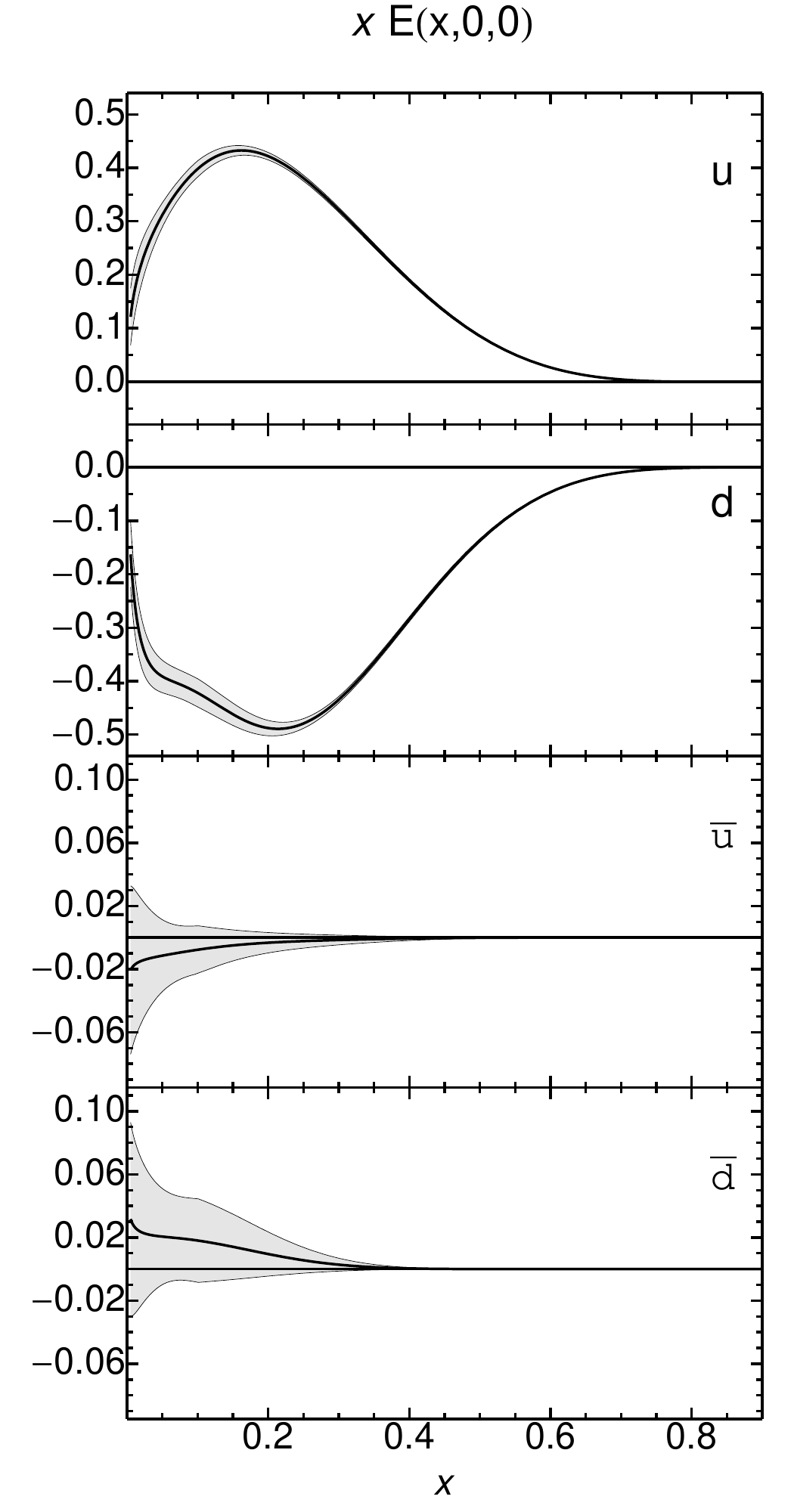}
\end{center}
\caption{The GPD $x E^a(x,0,0; Q_L^2)$ as a function of $x$ at the scale $Q_L^2 = 1$ GeV$^2$ for 
$a = u, d, \bar{u}, \bar{d}$ from top to bottom, respectively. The uncertainty bands are produced by statistical errors on the fit parameters.}
\label{fig:E}
\end{figure}

Our results for the Sivers function are comparable with other 
extractions~\cite{Vogelsang:2005cs,Anselmino:2008sga,Arnold:2008ap}. The results for the forward limit of the GPD $E$ are shown in Fig.~\ref{fig:E}; they also turn out qualitatively similar to available 
extractions~\cite{Guidal:2004nd,Diehl:2004cx,Goloskokov:2008ib,Goldstein:2010gu}.

\section{Results for (anti)quark total angular momenta}

\begin{widetext}

\begin{figure}[h]
\begin{center}
\includegraphics[width=19cm]{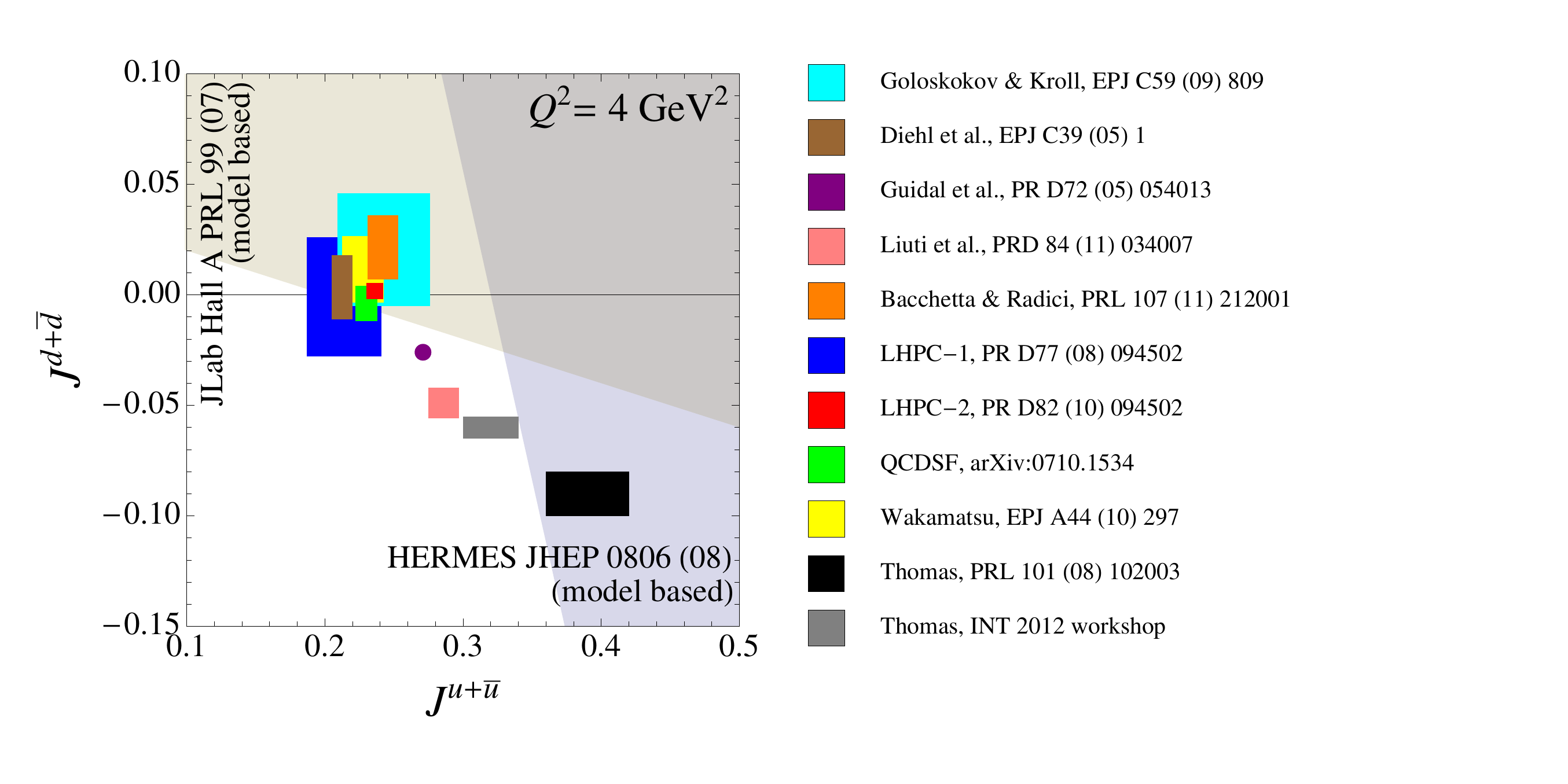}
\end{center}
\caption{The results of our determination of $J^{u+\bar{u}}$ and $J^{d+\bar{d}}$ compared with measurements of 
Refs.~\cite{Mazouz:2007aa,Airapetian:2008aa} and with other estimates based on GPD extractions from 
Refs.~\cite{Guidal:2004nd,Diehl:2004cx,Goloskokov:2008ib,Goldstein:2010gu}  (the values based on the parametrization~\cite{Goldstein:2010gu} are reported in Ref.~\cite{Taneja:2011sy}), on lattice 
simulations~\cite{Hagler:2007xi,Bratt:2010jn,Brommel:2007sb,Wakamatsu:2009gx}, and on some model calculations~\cite{Thomas:2008ga,Thomas:2012}.} 
\label{fig:JuJd}
\end{figure}

\end{widetext}

Using Eq.~\eqref{e:Jdef}, we can compute the total longitudinal angular momentum carried by each flavor $q$ and $\bar{q}$ at our initial scale $Q_L^2 = 1$ GeV$^2$. Using the standard evolution equations for the angular momentum (at leading order, with 3 flavors only, and $\Lambda_{\mathrm{QCD}} = 257$ MeV), we obtain the following results at $Q^2 = 4$ GeV$^2$:  

\begin{align*}
J^u &= 0.229 \pm 0.002^{+0.008}_{-0.012} ,   &J^{\bar{u}} &= 0.015 \pm 0.003^{+0.001}_{-0.000} ,  \\
J^d &= -0.007 \pm 0.003^{+0.020}_{-0.005} ,  &J^{\bar{d}} &= 0.022 \pm 0.005^{+0.001}_{-0.000} ,  \\
J^s &= 0.006^{+0.002}_{-0.006} ,  &J^{\bar{s}} &= 0.006^{+0.000}_{-0.005} . 
\end{align*} 
The first symmetric error is statistical and related to the errors on the fit parameters, while the second asymmetric error is theoretical and reflects the uncertainty introduced by other possible scenarios. In the present approach, we cannot include the (probably large) systematic error due to the rigidity of the functional form in
Eqs.~\eqref{e:gauss} and \eqref{e:lensing}. The bias induced by the choice of the functional form may affect in particular the determination of the sea quark angular momenta, since they are not directly constrained by
the values of the nucleon anomalous magnetic moments. Our present estimates (at $Q^2 = 4$ GeV$^2$) agree well with other estimates (see Fig.~\ref{fig:JuJd}), particularly with those ones based on the extraction of the GPD 
$E$~\cite{Guidal:2004nd,Diehl:2004cx,Goloskokov:2008ib} and on lattice 
simulations~\cite{Hagler:2007xi,Bratt:2010jn,Brommel:2007sb,Wakamatsu:2009gx}.  Our study indicates a total contribution to the nucleon spin from quarks and antiquarks of $0.271 \pm 0.007^{+0.032}_{-0.028}$, of which 85\% is carried by the up quark. 

Our approach can be used also to estimate the size of the total angular momentum carried by the gluons. In this case, we expect the lensing function to be different from that of the quarks. However, our extraction leaves little room for a nonzero gluon Sivers function, since the quark Sivers function already saturates the so-called Burkardt sum rule~\cite{Burkardt:2004ur}. If the Sivers function of the gluons is zero, our reasoning allows us to conclude
that $E^g$ is also zero, independent of the details of the lensing function. This would lead to a value of $J^g=0.215$ at 4 GeV$^2$. However, these considerations are strongly affected by the uncertainties on the
sea-quark Sivers functions outside the $x$ range where data exists. Direct measurements of the sea-quark and gluon Sivers functions are therefore highly necessary. 

At this point, we add a remark on the effect of TMD evolution on the Sivers function. The discussions in
Refs.~\cite{Aybat:2011ge,Aybat:2011ta,Anselmino:2012aa} suggest that the inclusion of TMD-evolution effects might lead to larger values of the Sivers function at the starting scale $Q_0^2$. If this were the case, we would need to compensate the effect by a smaller size of the lensing function in order to have an agreement with the anomalous magnetic moments. However, this will have a negligible net effect on the results for $J^a$.


\end{document}